\documentclass[a4paper,conference]{IEEEtran}
\IEEEoverridecommandlockouts

\usepackage{cite}
\usepackage{amsmath,amssymb,amsfonts}
\usepackage{algorithmic}
\usepackage{graphicx}
\usepackage{textcomp}
\usepackage[dvipsnames]{xcolor}
\def\BibTeX{{\rm B\kern-.05em{\sc i\kern-.025em b}\kern-.08em
    T\kern-.1667em\lower.7ex\hbox{E}\kern-.125emX}}

\usepackage[utf8]{inputenc}
\usepackage{hyperref}
\usepackage{url}
\usepackage[linesnumbered,ruled,vlined]{algorithm2e}
\usepackage{stfloats}
\usepackage{array}

\usepackage{draftwatermark}
\SetWatermarkText{Pre-print}
\SetWatermarkScale{1.3}
\SetWatermarkColor[gray]{0.85}

\begin{document}

\title{Grid-informed Sharing Coefficients in Renewable Energy Communities\\
\thanks{This research was funded by CETPartnership, the Clean Energy Transition Partnership 
under the 2022 joint call for research proposals, co-funded by the European Commission (GA N°101069750) 
and with the funding organizations detailed on \url{ https://cetpartnership.eu/funding-agencies-and-call-modules}, within the CoPRESS project framework.}
}

\author{\IEEEauthorblockN{Alireza Shooshtari}
\IEEEauthorblockA{\textit{Power system Group} \\
\textit{Catalonia Institute for Energy Research}\\
Barcelona, Spain \\
ashooshtari@irec.cat}
\and
\IEEEauthorblockN{Antonio Pepiciello}
\IEEEauthorblockA{\textit{Power system Group} \\
\textit{Catalonia Institute for Energy Research}\\
Barcelona, Spain \\
apepiciello@irec.cat}
\and
\IEEEauthorblockN{José Luis Domínguez-García}
\IEEEauthorblockA{\textit{Power system Group} \\
\textit{Catalonia Institute for Energy Research}\\
Barcelona, Spain \\
jldominguez@irec.cat}

}

% \author{\IEEEauthorblockN{Alireza Shooshtari, Antonio Pepiciello, José Luis Domínguez-García}
% \IEEEauthorblockA{
% Catalonia Institute for Energy Research  (IREC)\\
% Barcelona, Spain \\
% Email: ashooshtari, apepiciello, jldominguez@irec.cat}
% }

\maketitle

\begin{abstract}
The role of energy communities in grid operations is highly dependent on the spatial distribution of their participants. In particular, when local energy producers and consumers are concentrated in different feeders, economic incentives from energy communities have the potential to affect local grid congestion.
To address this challenge, we propose a feeder-aware allocation strategy that reflects grid topology in energy sharing.
This strategy prioritizes energy sharing within the same feeder, thus incentivizing local generation-demand balance and improving grid operation. Different sharing coefficients are tested, such as equal, proportional, and rank-based, in both static and dynamic formulations.
The proposed strategy is tested on data from a real energy community, whose participants are assumed to be distributed across four feeders.
The analysis is carried out from the perspectives of the community as a whole, individual feeders, and single participants.
Simulation results show that the feeder-aware strategy, in addition to promoting local energy balance, leads to higher and more stable revenues for most participants.
\end{abstract}

\IEEEpubid{\begin{minipage}{\textwidth}\vspace{1.5cm}
979-8-3315-9515-9/25/\$31.00~\copyright~2025 IEEE
\end{minipage}}


\section{Introduction}
Decarbonization and decentralization of power systems have become increasingly important due to concerns about climate change and energy security, while developments in the grid integration of distributed energy resources have made this transition more feasible. In this regard, the European Union has established new rules for Energy Communities (ECs) through the “Clean Energy Package” \cite{ec2019}, which was later clarified through two directives: the EU Renewable Energy Directive 2018/2001 (RED II) \cite{eu2018} and the EU Internal Electricity Market Directive 2019/944 (IEMD) \cite{eu2019}. Within this framework, alongside the introduction of active consumers, collective self-consumption (CSC)  and renewable energy community (REC) regulations were introduced as part of the RED II and IEMD directives.

An energy community enables consumers to actively participate in the generation and consumption chain by owning distributed generation units, transforming them from passive consumers into prosumers. Although the economic benefits of collective self-consumption in ECs are evident, there are additional advantages, including balancing local supply and demand \cite{herenvcic2019effects}, improving environmental conditions, enhancing energy security \cite{llera2023effect}, and reducing grid losses \cite{fernandez2021profitability}. Within an energy community, energy producers benefit from multiple types of transactions, as the framework supports both energy exchange with the power grid and peer-to-peer energy trading, thereby increasing income and reducing costs \cite{tostado2023optimal}.

One of the key factors in optimally managing ECs and evaluating participants' profitability is the allocation of surplus energy among members. This allocation is defined by sharing coefficients, which are calculated based on production and consumption data sent by the distribution system operator (DSO) to the energy community management and then returned to the DSO for integration into the metering and settlement process. The way these sharing coefficients are calculated can significantly influence the distribution of benefits among participants.

In recent years, extensive research has been conducted to examine the fairness of cost-benefit distribution in ECs, their real-world validation, economic advantages, and performance under different sharing coefficient schemes. According to analyses based on Jain's index and other fairness metrics, the Virtual Net-Billing method, which divides the benefits of renewable energy among coalition members according to their energy contributions and the price difference between imported and exported electricity, performs the best when evaluating the fairness of three different energy-sharing strategies within ECs \cite{gjorgievski2022fairness}. Similarly, three redistribution mechanisms are proposed in \cite{fina2024energy} to enhance the fairness of network cost sharing within ECs, with each method aiming to balance individual incentives and collective equity. In the context of fair cost and benefit allocation, shared solar generation in multi-family buildings in Sydney is examined \cite{roberts2022efficient}. The study emphasizes the challenge of designing tariff structures that ensure both investment recovery and equitable benefit distribution among all residents in an embedded network.

To validate energy community models in practice, several studies have used real data and existing ECs. For instance, the impact of new tariffs and the use of variable sharing coefficients were demonstrated using data from a Spanish energy community in \cite{manso2021local}. In \cite{fina2023measures}, a case from Austria was analyzed with the aim of enhancing the diffusion and integration of ECs through the examination of stakeholder relationships and regulatory barriers. Additionally, case studies from Tolosa and the Croatian city of Križevci, showed that the appropriate configuration of ECs, along with regulatory support and effective pricing mechanisms, can enhance self-sufficiency as well as economic, social, and environmental benefits \cite{lopez2023photovoltaic, herenvcic2022automated}.

Sharing coefficient schemes and their impacts on the technical performance and profitability of ECs have been the focus of considerable research. In \cite{fina2022simulation}, a comparison was made between static and dynamic sharing coefficients, and it was found that the use of dynamic coefficients improved local energy consumption and economic outcomes for all participants. Similarly, different sharing coefficients were evaluated in a Portuguese energy community in \cite{queiroz2023assessment}, and dynamic coefficients were found to enhance overall performance. In addition, it has been shown that implementing self-consumption before sharing energy within the community reduces the profit gap among participants. In another Spanish case study, the application of dynamic distribution coefficients based on consumption was confirmed to significantly improve economic performance. The study considered fixed coefficients on a yearly and monthly basis, as well as dynamic coefficients that were updated hourly in proportion to each participant’s energy consumption. While investment-proportionate coefficients are considered fairer from an equity point of view, consumption-based coefficients have been identified as more effective in terms of efficiency \cite{llera2023effect}. Lastly, in \cite{gianaroli2024development}, four dynamic allocation algorithms for energy sharing in renewable ECs were developed and evaluated. The limitations of widely used consumption-proportional methods were highlighted, and time-sensitive, behavior-aware criteria, such as synchronism and sharing trends, were suggested to enhance fairness, efficiency, and sustainability.

Although extensive research has explored energy community management and the use of various sharing coefficients, to the best of our knowledge, no previous study has considered the power grid configuration underlying the energy community. For this reason, our grid-informed approach examines prioritization at the feeder level within a framework we call feeder-aware. This strategy prioritizes participants connected to the same feeder for energy allocation before distributing energy across the broader energy community. This prioritization can avoid unwanted grid congestion caused by eventual imbalances in generation and demand among the participants in the EC. To evaluate the effectiveness of this approach, we apply three types of sharing coefficients: equal, proportional, and rank-based, under both dynamic and static calculations. For a more comprehensive analysis, this strategy is compared to a baseline scenario without energy community implementation and also feeder prioritization, which we refer to as feeder-agnostic.

The paper is structured as follows: Section 2 details the methodology, including allocation strategies, sharing coefficients, the dataset, and the tariff structure. Section 3 presents the results and comparison, and Section 4 concludes with key insights and future directions.
\section{Methodology}

\subsection{Allocation strategy}
The purpose of the allocation strategy is to distribute the community's energy surplus among all participants. The main objective of designing an effective allocation strategy is to reduce surplus energy exported to the grid and to ensure fair compensation for all members.

In this study, it is assumed that before any allocation, every prosumer satisfies their demand with self-consumption. 
The surplus \( S_{i,t} \) and demand \( D_{i,t} \) of each participant $i$ at time $t$ are computed as in (\ref{eq:surplus}) and (\ref{eq:need}), based on behind-the-meter generation $G_{i,t}$ and consumption $C_{i,t}$ data, assumed to be available.
\begin{equation}
S_{i,t} = \max \left( G_{i,t} - C_{i,t},\ 0 \right)\label{eq:surplus}
\end{equation}
\begin{equation}
D_{i,t} = \max \left( C_{i,t} - G_{i,t},\ 0 \right)\label{eq:need}
\end{equation}

According to the sharing coefficients, surplus energy must be allocated to participants within the EC. If the amount of allocated energy to each participant, {$A_i$}, exceeds their demand at each time interval, the remaining part will be exported to the grid. 

In this paper, we introduce a feeder-aware strategy that gives priority to participants connected to the same feeder. This contrasts with the feeder-agnostic approach, where surplus energy is shared among all participants, regardless of their location. This reduces unnecessary energy transfers between feeders, which can help alleviate congestion on transformers and inter-feeder links. Ultimately, the feeder-aware strategy aligns financial incentives with physical grid constraints, offering technical benefits such as improved local balancing and potentially reduced distribution losses due to shorter energy delivery paths.

Under both feeder-aware and feeder-agnostic allocation strategies, a participant's allocated energy can come from prosumers either inside the same feeder or from others. In order to distinguish the source of allocated energy, $A_i$ is defined as follows:
\begin{equation}
A_i=A_i^{\text{same}}+A_i^{\text{other}}
\end{equation}
where $A_i^{\text{same}}$ represents the energy allocated to the participant $i$ from prosumers within the same feeder, and $A_i^{\text{other}}$, from other feeders. In the feeder-aware strategy, surplus energy is first allocated among consumers within the same feeder. If some consumers still need more energy and there is surplus remaining after this local sharing, they can receive additional surplus from other feeders.

Both static and dynamic methods can be used to calculate sharing coefficients. 
The key difference between static and dynamic sharing coefficients is that in the former case, energy is allocated among participants based on calculations carried out at the beginning of the year. However, with the dynamic method, sharing coefficients must be recalculated at each time interval. The main advantage of the static allocation is its simplicity: the participants within the EC know their sharing coefficients beforehand and can estimate the economic benefit of being part of the community accurately. On the other hand, dynamic allocation introduces uncertainty in the yearly revenues and costs for each participant, but it enables a more efficient allocation of the surplus energy, reducing the total energy sold to the grid.

The algorithm for the allocation of energy at each time interval has been provided in Algorithm~\ref{alg:allocation}.

\begin{algorithm}[!htbp]
\DontPrintSemicolon
\SetKwInOut{Input}{Input}
\SetKwInOut{Output}{Output}

\Input{Generation and consumption profiles, feeder ID}
\Output{Energy allocation from the community to all participants}

\ForEach{participant $i$}{
    Compute energy surplus or demand\;
}

\If{strategy == \texttt{feeder-aware}}{
    \ForEach{feeder $f$}{
        Calculate total surplus within feeder $f$\;
        Allocate surplus among participants in feeder $f$ using sharing coefficients\;
    }
    Allocate remaining surplus among all participants in the community\;
}
\ElseIf{strategy == \texttt{feeder-agnostic}}{
    Calculate total surplus in the energy community\;
    Allocate total surplus among all participants in the community\;
}

\caption{Energy Allocation: Feeder-Aware vs. Feeder-Agnostic}
\label{alg:allocation}
\end{algorithm}

\subsection{Sharing Coefficients}
Three types of sharing coefficients are considered: equal, proportional to energy demand, and rank-based, prioritizing participants with the lowest demand.

To calculate equal sharing coefficients using the static method, the total surplus is divided by the number of participants (\( {N_{\text{total}}}\)) in every feeder, when choosing the feeder-aware allocation strategy, and by the total number of EC participants, in both the feeder-aware's global allocation phase and the feeder-agnostic one. However, in the dynamic method,  surplus energy is divided by the number of participants who need energy (\({N_{\text{need},t}}\)) at each time interval, meaning that participants with $D_i=0$ are excluded from the allocation. This ensures that surplus energy is allocated only to participants with unmet demand, resulting in a more efficient energy allocation.

Proportional sharing coefficients are based on each participant’s share of the total demand. In the static method, the coefficients are calculated at the beginning of the year by dividing each participant’s total yearly demand (\({D_{i,\text{year}}}\)) by the total yearly demand of their feeder and the entire energy community. These coefficients remain constant throughout the year and are used for energy allocation in both feeder-aware and feeder-agnostic approaches. 

In the dynamic method, however, the sharing coefficients change over time. They are recalculated at every time interval, based on each participant’s share of the feeder’s and the EC’s total demand at that moment. For more details, the calculation of equal and proportional sharing coefficients under the static and dynamic methods is provided in Table \ref{tab:sharing_coefficients}.
\begin{table}[!b]
\footnotesize
\centering
\caption{Sharing coefficient calculation under static and dynamic methods.}
\begin{tabular}{|c|c|c|}
\hline
\textbf{Sharing Coefficient Type} & \textbf{Static Method} & \textbf{Dynamic Method} \\
\hline
Equal & 
$\alpha_{i} = \frac{1}{N_{\text{total}}}$ & 
$\alpha_{i,t} = \frac{1}{N_{\text{need},t}}$ \\
\hline
Proportional & 
$\alpha_{i} = \frac{D_{i,\text{year}}}{\sum_j D_{j,\text{year}}}$ & 
$\alpha_{i,t} = \frac{D_{i,t}}{\sum_j D_{j,t}}$ \\
\hline
\end{tabular}
\label{tab:sharing_coefficients}
\end{table}

Lastly, to calculate rank-based sharing coefficients, participants are ranked according to their energy demand, beginning with the one who uses the least energy. The second lowest is taken into account next, and so on. According to their total annual demand, we rank them once at the start of the year using the static method. In the dynamic method, the ranking changes over time, and it’s updated at each time step based on their current demand.
\subsection{Data \& Pricing}
This work is part of a larger project on ECs in Switzerland. To evaluate our proposed strategy, Swiss electricity prices were applied to a comprehensive energy-related dataset collected from 
a publicly available dataset in the literature \cite{trivedi2024comprehensive}, as the Swiss project data are confidential. The dataset includes per-household energy generation and consumption data at a one-minute time interval. Due to data availability, our analysis covers the period from January 1st to December 31st, 2020.  In order to mimic the characteristics of a Swiss EC, we randomly selected 15 buildings from the available dataset and assigned them to four feeders, as shown in Table \ref{tab:feeder_assignments}. Within this set, households 6, 8, 9, 12, 14, and 15 are consumers, while the remaining households are prosumers.

\begin{table}[!b]
\centering
\caption{Feeder assignments of selected buildings}
\label{tab:feeder_assignments}
\begin{tabular}{|c|c|}
\hline
\textbf{Feeder} & \textbf{Building IDs} \\
\hline
Feeder 1 & H1, H9, H10, H11, H13, H14 \\
\hline
Feeder 2 & H2, H3 \\
\hline
Feeder 3 & H4, H5, H6, H12 \\
\hline
Feeder 4 & H7, H8, H15 \\
\hline
\end{tabular}
\end{table}

Typical household electricity prices from a Swiss municipality were used for the analysis. The feed-in price for sending electricity back to the grid is 0.08 EUR/kWh. The price for consuming electricity from the grid, considering a flat rate that represents a constant price per kilowatt-hour regardless of the time of day or week, is 0.2562 EUR/kWh, composed of: \begin{itemize} \item 0.1137 EUR for energy, \item 0.1092 EUR for network use, \item 0.0333 EUR for other charges. \end{itemize}
In the energy community, both the selling price for prosumers and the buying energy price for consumers are set at 0.1137 EUR/kWh. If the allocated energy comes from prosumers on the same feeder, the consumer receives a 40\% discount on the network usage charge. If the energy comes from a different feeder, the discount on the network charge is 20\%.

\section{Results}
The performance of the energy community is compared under both feeder-aware and feeder-agnostic strategies, using the three previously defined sharing coefficients. The results are compared to a baseline where no energy community is present, and each participant operates individually. To provide a comprehensive understanding of the technical and economic impacts, the analysis is carried out from three perspectives: the energy community as a whole, individual feeders, and individual participants. All evaluations were carried out in Python, and the implementation is publicly available on GitHub \cite{github2025}.

From the energy community’s perspective, as shown in Table \ref{tab:comparison}, dynamic sharing coefficients were more effective in maximizing the amount of energy shared within the community and minimizing energy exports from and imports to the community. This indicated an increase in the self-sufficiency of EC. The identical values observed in the dynamic proportional, dynamic rank-based, and static rank-based cases result from the fact that, in these cases, all available energy is fully allocated within the energy community.

\begin{table*}[!b]
\caption{Energy sharing performance of the EC under feeder-aware and feeder-agnostic strategies.}
\label{tab:comparison}
\centering

% ----- Feeder-aware Subtable -----
\vspace{1ex}
\textbf{(a) Feeder-aware} \\[0.5ex]
\makebox[\textwidth][c]{%
\begin{tabular}{
    |>{\centering\arraybackslash}p{4cm}
    |>{\centering\arraybackslash}p{3.5cm}
    |>{\centering\arraybackslash}p{3.5cm}
    |>{\centering\arraybackslash}p{3.5cm}|
}
\hline
\textbf{Sharing Coeff.} & \textbf{Imported Energy [MWh]} & \textbf{Shared Energy [MWh]} & \textbf{Exported Energy [MWh]} \\
\hline
Dynamic equal        & 59.212 & 13.433 & 3.287 \\
\hline
Static equal         & 61.144 & 11.501 & 5.219 \\
\hline
Dynamic proportional & 58.702 & 13.942 & 2.777 \\
\hline
Static proportional  & 61.930 & 10.715 & 6.005 \\
\hline
Dynamic rank-based   & 58.702 & 13.942 & 2.777 \\
\hline\
Static rank-based    & 58.702 & 13.942 & 2.777 \\
\hline
\end{tabular}
}

% ----- Spacing between subtables -----
\vspace{2ex}

% ----- Feeder-agnostic Subtable -----
\textbf{(b) Feeder-agnostic} \\[0.5ex]
\makebox[\textwidth][c]{%
\begin{tabular}{
    |>{\centering\arraybackslash}p{4cm}
    |>{\centering\arraybackslash}p{3.5cm}
    |>{\centering\arraybackslash}p{3.5cm}
    |>{\centering\arraybackslash}p{3.5cm}|
}
\hline
\textbf{Sharing Coeff.} & \textbf{Imported Energy [MWh]} & \textbf{Shared Energy [MWh]} & \textbf{Exported Energy [MWh]} \\
\hline
Dynamic equal        & 60.329 & 12.316 & 4.404 \\
\hline
Static equal         & 61.937 & 10.707 & 6.012 \\
\hline
Dynamic proportional & 58.702 & 13.942 & 2.777 \\
\hline
Static proportional  & 61.733 & 10.912 & 5.808 \\
\hline
Dynamic rank-based   & 58.702 & 13.942 & 2.777 \\
\hline
Static rank-based    & 58.702 & 13.942 & 2.777 \\
\hline
\end{tabular}
}
\end{table*}

\begin{figure}[!t]
\includegraphics[width=\columnwidth]{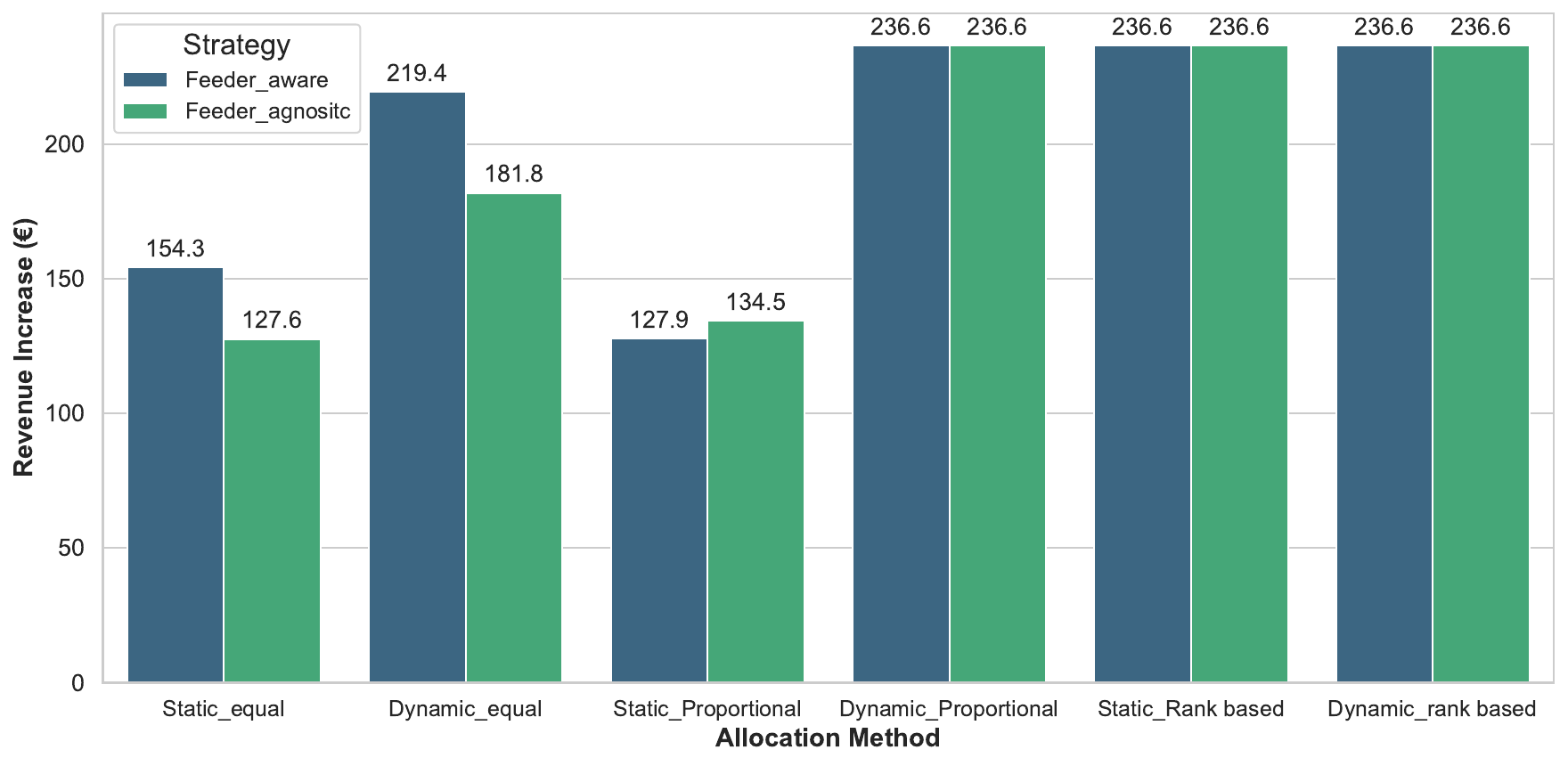}
\caption{Comparison of revenue increase in the EC under feeder-aware and feeder-agnostic allocation strategies.}
\label{fig:revenue}
\end{figure}

Also, the increase in revenue of the energy community is illustrated in Figure \ref{fig:revenue}, where both feeder-aware and feeder-agnostic strategies demonstrate a significant increase in revenue from selling energy within the community compared to the case without an energy community, particularly when dynamic sharing coefficients are used. The lower revenue increase of the feeder-aware strategy with static proportional sharing coefficients is due to a mismatch between participants' fixed shares, based on annual demand, and their real-time energy needs. In feeders where this mismatch occurs, a significant portion of surplus energy remains unused during the feeder-level allocation. When this surplus is subsequently reallocated at the community level, it follows the static proportions of all EC participants, even though many have already satisfied their demand through their feeder allocation. As a result, much of the surplus energy remains unused and is exported to the grid, leading to a reduction in participants' revenue. 

\begin{figure}[!t]
\includegraphics[width=\columnwidth]{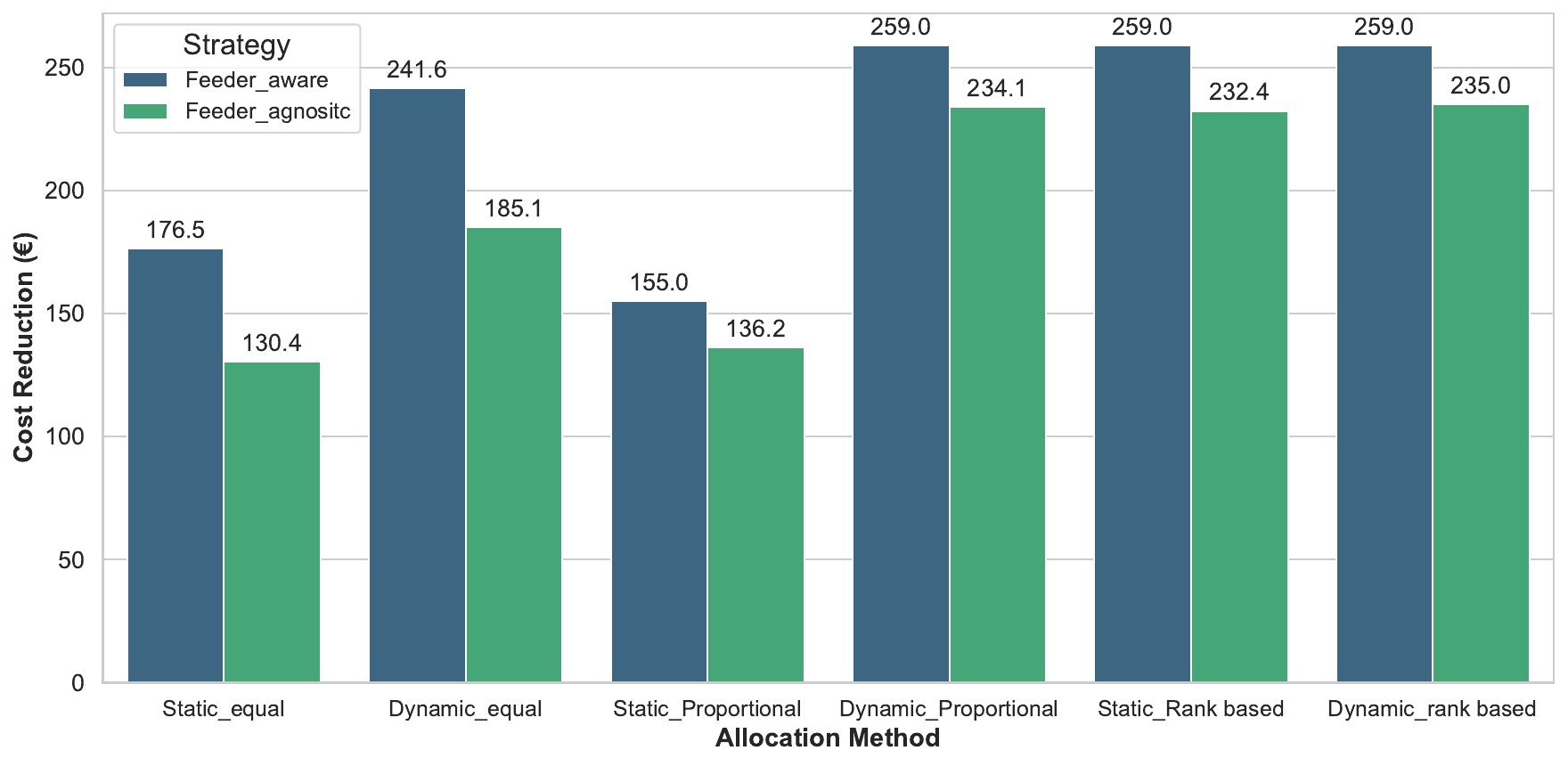}
\caption{Comparison of cost reduction in the EC under feeder-aware and feeder-agnostic allocation strategies.}
\label{fig:cost}
\end{figure}

Regarding the reduction in energy purchase costs, as shown in Figure \ref{fig:cost}, the feeder-aware strategy outperforms the feeder-agnostic approach across all cases. Furthermore, it can be highlighted that the choice of feeder-aware strategy strongly affects the EC, compared to feeder-agnostic one, in the static and dynamic equal cases. This is mainly due to the way surplus energy is allocated in the Feeder-aware strategy. In this approach, surplus energy is shared locally within each feeder, so it’s only distributed among the participants connected to the same feeder. As a result, in feeders where there’s both a lot of surplus energy and high demand, each participant can get a bigger share because the group is smaller. This local sharing makes better use of the available energy and helps reduce costs more effectively. On the other hand, the Feeder-agnostic strategy aggregates the surplus energy from all feeders and distributes it equally among all community members, regardless of their individual feeders. This can lead to some of the surplus energy going unused and people with higher needs not getting enough.

These results highlight the potential to incentivize participants to join the energy community, as they can not only reduce their electricity costs but also generate revenue if they produce excess energy. 

\begin{figure}[!t]
\includegraphics[width=\columnwidth]{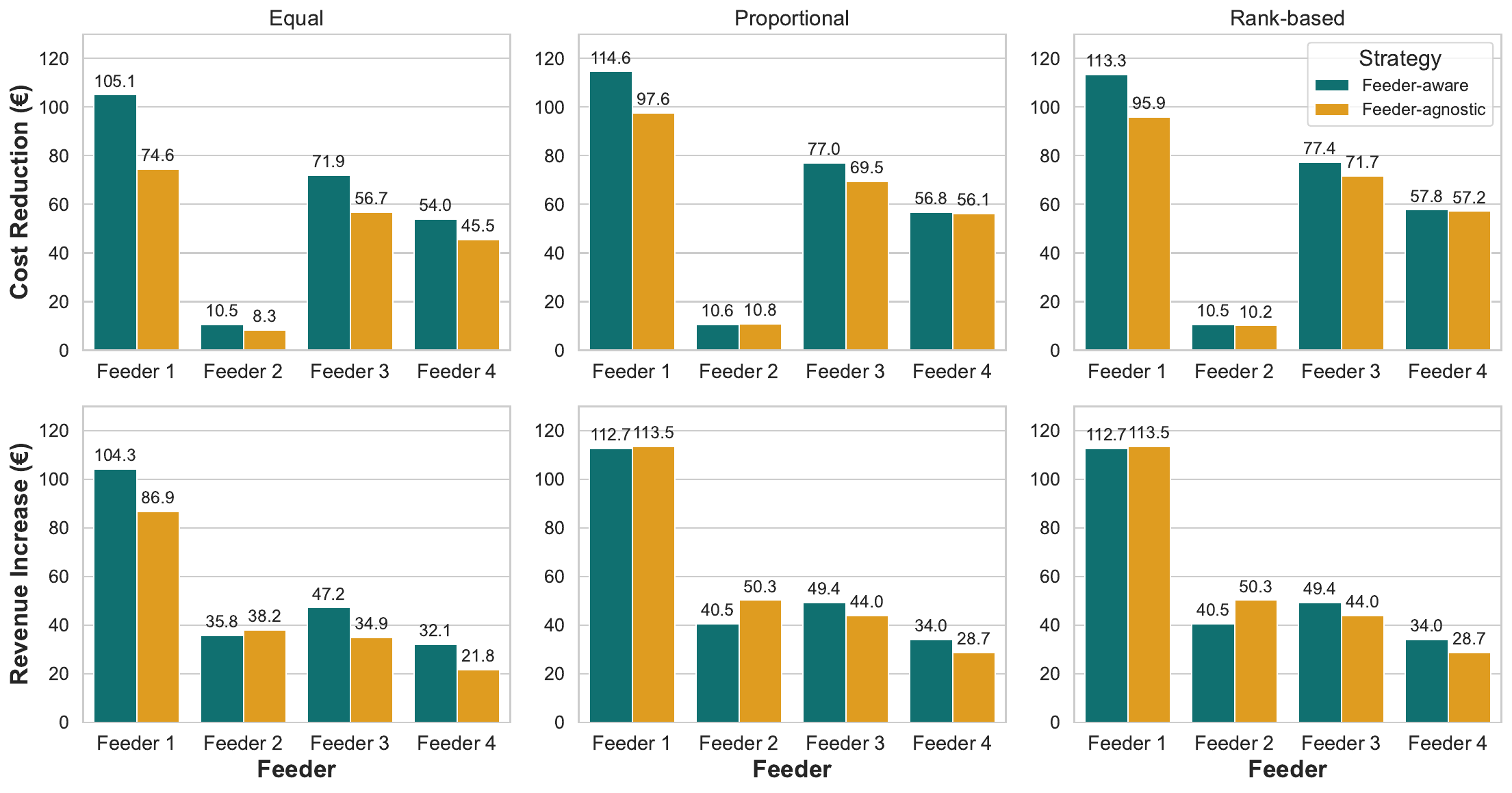}
\caption{Distribution of financial benefits among feeders using dynamic sharing coefficients.}
\label{fig:feeder-dynamic}
\end{figure}

\begin{figure}[!t]
\includegraphics[width=\columnwidth]{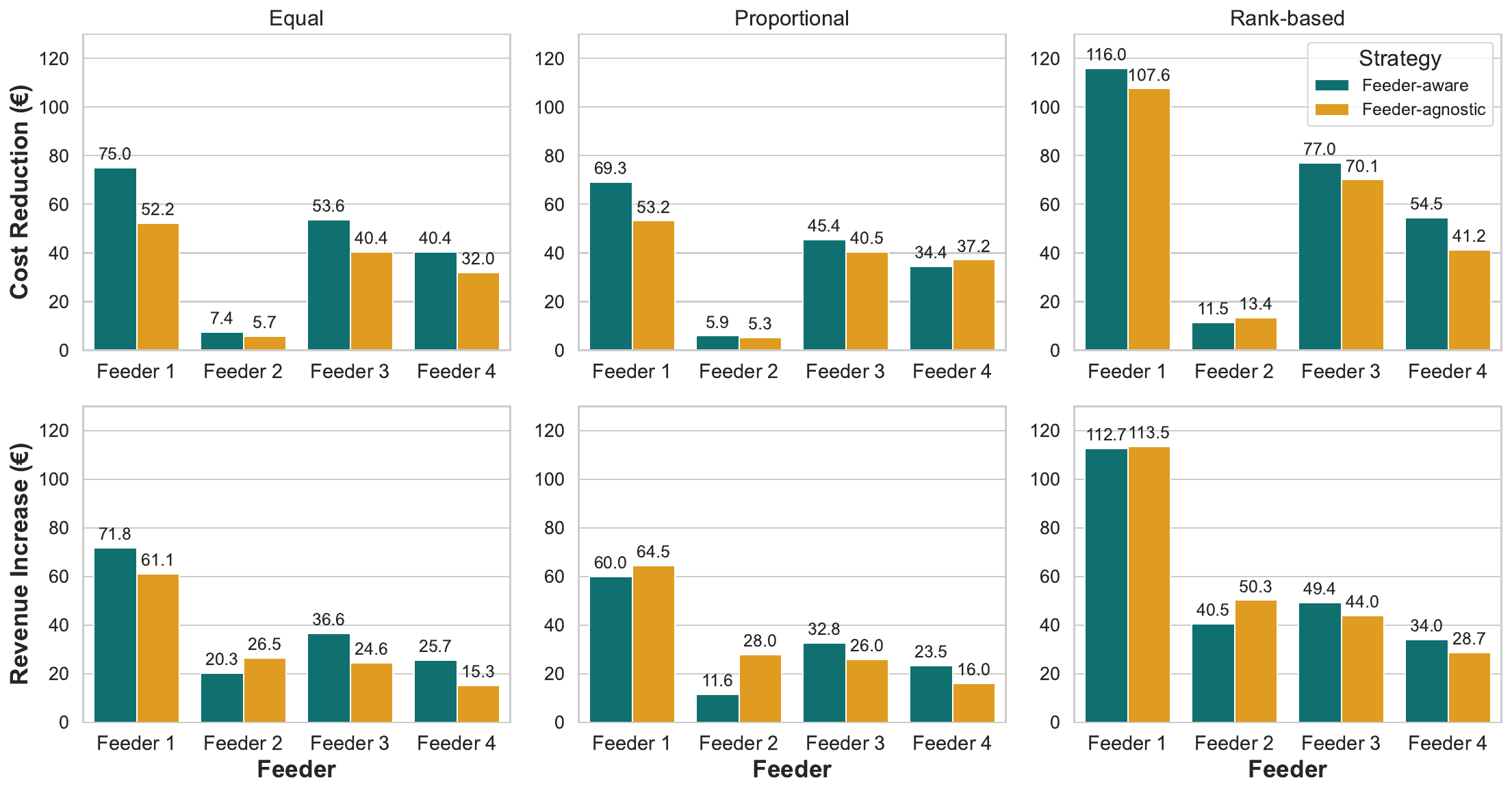}
\caption{Distribution of financial benefits among feeders using static sharing coefficients.}
\label{fig:feeder-static}
\end{figure}

From the feeder point of view, the financial benefit distribution across all feeders with the application of dynamic and static sharing coefficients is shown in Figures \ref{fig:feeder-dynamic} and \ref{fig:feeder-static}, respectively. Although in most cases, and particularly in terms of cost reduction, the feeder-aware approach provides higher financial benefits, there are some differences when it comes to revenue increase. Some feeders benefit more from selling energy under the feeder-aware strategy, while others do so under the feeder-agnostic strategy. It is more beneficial for participants in Feeder 2 and Feeder 1 to allocate surplus energy using the feeder-agnostic approach. This is because the density of producers is higher in these feeders, making it more advantageous for them to sell their surplus energy to other feeders. With the feeder-aware strategy, their participation is limited to within their own feeder, which can reduce their revenue potential.

\begin{figure}[!t]
\includegraphics[width=\columnwidth]{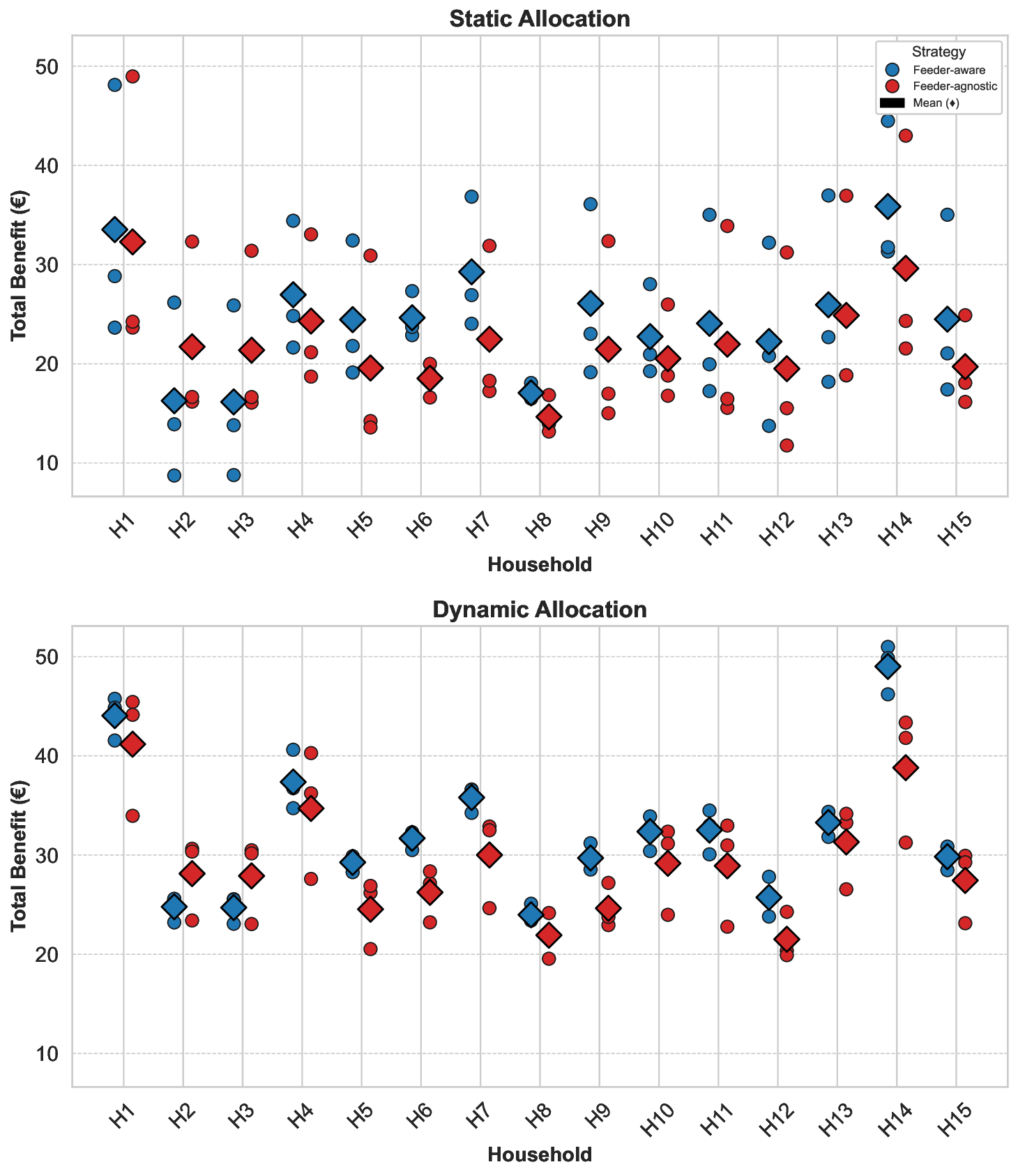}
\caption{Distribution of financial benefits among participants.}
\label{fig:boxplot}
\end{figure}

Finally, from the participants’ perspective, the total financial benefit achieved by each household when using both static and dynamic sharing coefficients is illustrated in Figure \ref{fig:boxplot}. The total benefit is defined as the sum of cost reduction and revenue increase per household. Each point corresponds to one of the three sharing methods (equal, proportional, and rank-based) under either the feeder-aware or feeder-agnostic strategy. Diamonds represent the average of these three values. The results indicate that for the majority of the participants, the average total benefit is higher under the feeder-aware strategy, both with static and dynamic sharing coefficients. This means that using a feeder-aware strategy enables participants to achieve higher financial benefits from participation in the energy community.
In case of Households 2 and 3, who are the only members connected to Feeder 2 and both are producers with energy surplus, the average total benefit is higher under the feeder-agnostic strategy. This is because, in the feeder-agnostic strategy, they are able to sell this surplus energy across the entire community, maximizing their potential financial benefit. In contrast, under the feeder-aware strategy, participants’ energy needs are prioritized within their own feeders. As a result, some of the demand in other feeders is already satisfied locally, which limits Households 2 and 3's ability to sell their surplus energy beyond their own feeder. This restriction reduces their financial benefit compared to the feeder-agnostic approach.

Furthermore, the feeder-aware strategy with dynamic allocation exhibits lower variance in the total benefits achieved across the three sharing coefficient methods. This suggests that households experience less uncertainty regarding their expected financial outcomes under the feeder-aware approach. Such predictability may enhance participants’ trust and confidence in the energy community.

\section{Conclusion}
This paper proposed a feeder-aware strategy that prioritizes energy sharing among energy community participants connected to the same feeder. This approach aims to incentivize local supply-demand balance and mitigate the risk of grid congestion or other technical issues associated with the economic incentives of the EC.
The effects of the proposed feeder-aware strategy were assessed considering data from an energy community whose participants are distributed across four feeders. The feeder-aware strategy is compared with a feeder-agnostic one, considering the scenario without energy community as the benchmark.
Different sharing coefficients were evaluated for each strategy: equal, proportional, and rank-based ones, for both static and dynamic formulations.
Results show that from a community-wide perspective, the feeder-aware strategy leads to higher cost reduction compared to the feeder-agnostic case.
However, this varied across feeders, with more balanced feeders benefitting the most.
Among the tested sharing coefficients, dynamic proportional, static rank-based, and dynamic rank-based ones resulted in the highest levels of shared energy and participant revenues.
At the participant level, the feeder-aware strategy appears to mitigate the revenue uncertainty related to dynamic coefficients, especially for participants in balanced feeders.
Future work could focus on investigating how the spatial distribution of consumers and producers across the energy community affect power flow, congestion and grid operation. Furthermore, the grid-aware strategy can be enhanced by incorporating electrical variables, such as line currents or node voltages, to enhance the EC's role in supporting reliable grid operation.

% \section*{Acknowledgment}

% This research was funded by CETPartnership, the Clean Energy Transition Partnership 
% under the 2022 joint call for research proposals, co-funded by the European Commission (GA N°101069750) 
% and with the funding organizations detailed on \url{ https://cetpartnership.eu/funding-agencies-and-call-modules}, within the CoPRESS project framework.

% \bibliographystyle{IEEEtran}
% \bibliography{references}

\bibliographystyle{IEEEtran}

\vspace{12pt}
\color{red}
\end{document}